\documentclass[12pt]{article}

\usepackage{mathtools}
\usepackage{booktabs}
\usepackage[english]{babel} % English language/hyphenation
\usepackage{amsmath,amsfonts,amsthm}
\usepackage{ amssymb }
\usepackage{enumitem}
\usepackage{graphicx}
\usepackage{array}
\usepackage[toc,page]{appendix}

\usepackage{booktabs} % Horizontal rules in tables
\usepackage{natbib}
\usepackage{setspace}
\usepackage{tabularx}
\usepackage{subcaption}
\usepackage[font = small,labelfont=bf,textfont=it]{caption} % Custom captions under/above floats in tables or figures
\usepackage{footnote}
\usepackage{multirow}
\makeatletter
\newcommand{\distas}[1]{\mathbin{\overset{#1}{\kern\z@\sim}}}%
\newsavebox{\mybox}\newsavebox{\mysim}

\theoremstyle{definition}

\newcommand{\distras}[1]{%
  \savebox{\mybox}{\hbox{\kern3pt$\scriptstyle#1$\kern3pt}}%
  \savebox{\mysim}{\hbox{$\sim$}}%
  \mathbin{\overset{#1}{\kern\z@\resizebox{\wd\mybox}{\ht\mysim}{$\sim$}}}%
}
\newcolumntype{C}[1]{>{\centering\let\newline\\\arraybackslash\hspace{0pt}}m{#1}}

\usepackage{color}

\newcommand{\blind}{1}

\begin{document}

\def\spacingset#1{\renewcommand{\baselinestretch}%
{#1}\small\normalsize} \spacingset{1.3}

%%%%%%%%%%%%%%%%%%%%%%%%%%%%%%%%%%%%%%%%%%%%%%%%%%%%%%%%%%%%%%%%%%%%%%%%%%%%%%

\if1\blind
{
 
 \centering{\bf\Large Advancing Inverse Scattering with Surrogate Modeling and Bayesian Inference for Functional Inputs}\\
  \vspace{0.2in}
  \centering{Chih-Li Sung$^{a}$, Yao Song$^{b}$, Ying Hung$^{b}$\footnote{The authors gratefully acknowledge funding from NSF DMS-2113407, DMS-2107891, and  CCF-1934924.}\vspace{0.2in}\\
        $^{a}$Michigan State University\\
    $^{b}$Rutgers, the State University of New Jersey\\}
    \date{\vspace{-7ex}}
  %\maketitle
} \fi

\if0\blind
{
  \bigskip
  \bigskip
  \bigskip
    \begin{center}
    {\LARGE\bf Advancing Inverse Scattering with Surrogate Modeling and Bayesian Inference for Functional Inputs
    %A Bayesian Framework for Inverse Scattering Problems with Functional Inputs
    %Bayesian Inverse Problems for Functional Inputs with Applications to Inverse Scattering Problems
    }
\end{center}
  \medskip
} \fi

\bigskip
\begin{abstract}

%Inverse scattering aims to infer information about a hidden object given the received scattered waves which is of interest in science, engineering, and medical studies. 
Inverse scattering aims to infer information about a hidden object by using the received scattered waves and training data collected from forward mathematical models. Recent advances in computing have led to increasing attention towards functional inverse inference, which can reveal more detailed properties of a hidden object. However, rigorous studies on functional inverse, including the reconstruction of the functional input and quantification of uncertainty, remain scarce. Motivated by an inverse scattering problem where the objective is to infer the functional input representing the refractive index of a bounded scatterer, a new Bayesian framework is proposed. It contains a surrogate model that takes into account the functional inputs directly through kernel functions, and a Bayesian procedure that infers functional inputs through the posterior distribution.  
%Inverse scattering problems are ubiquitous in various fields of science and engineering, where the goal is to estimate the properties or characteristics of a hidden object or medium from scattered wave data. In this study, we focus on an inverse scattering problem where the refractive index of a bounded scatterer serves as the functional input of interest. 
%To address the challenge of estimating functional inputs in inverse scattering problems efficiently and accurately, we propose a Bayesian framework that leverages a novel surrogate model with functional inputs. Our approach utilizes a Gaussian process prior for the inverse solution without the need of finite basis expansion, overcoming limitations of traditional methods. 
%Furthermore, the Bayesian framework is extended to integrate multi-fidelity simulations which leads to enhancement on the accuracy in functional recovery. 
%Additionally, we incorporate multifidelity simulations by combining high-fidelity and low-fidelity simulations with functional inputs in our surrogate model to enhance predictions for higher-fidelity simulations. 
%This enables us to achieve accurate and fast estimations of the functional inputs, making it suitable for our case study. 
Furthermore, the proposed Bayesian framework is extended to reconstruct functional inverse by integrating multi-fidelity simulations, including a high-fidelity simulator solved by finite element methods and a low-fidelity simulator called the Born approximation. When compared with existing alternatives developed by finite basis expansion, the proposed method provides more accurate functional recoveries with smaller prediction variations.
%The proposed approach shows promising results in solving inverse scattering problems with functional inputs, opening up new possibilities for advanced imaging and characterization techniques and highlighting the potential of Bayesian surrogate modeling in the field of inverse scattering.
\end{abstract}

\noindent%
\vfill

\newpage
\spacingset{1.45} % DON'T change the spacing!

\section{Introduction}\label{sec:intro}

Scattering problems refer to the scattering of waves which describe how waves interact with objects. 
%Scattering problems refer to the phenomenon of waves interacting with objects or obstacles in their path. 
After using waves to probe a hidden object, inverse scattering aims to infer information about the hidden object given the received response waves \citep{kaipio2006statistical,CCHbook}. There are broad applications of inverse scattering in various scientific fields, such as medical imaging, non-destructive testing, remote sensing, and radar imaging, among others. 
%For example in .... Various applications can be found in radar scattering and ultrasound imagining (reference), inverse scattering is applied to determine the index of refraction of a bounded medium from the respond far field pattern. 
For instance, in electrical impedance tomography (EIT), inverse scattering is used to infer the electric conductivity  \citep{dbar2020}, which reveals crucial medical information for the diagnosis of pulmonary embolism, detection of tumors in the chest area, and the diagnosis and distinction of ischemic and hemorrhagic stroke. Another important application is the computerized tomography (CT), which is widely used in medical studies for interior reconstruction that creates detailed cross-sectional images of the human body \citep{courdurier2008solving,li2019learning}.

%To infer properties of the hidden objects given measurements of the received waves, mathematical models, also known as forward solves, are developed to simulate the response waves for different properties of the objects (references). There are various types of forward solvers. For examples, nonlinear forward solvers are accurate but often computationally intensive which limits their applicability. On the other hand, a popular alternative is to approximate the nonlinear solvers by a linearized version of the problems, such as the use of the Born approximation.  These approximations are computationally efficient but potentially less accurate as compared to the nonlinear solvers. The goal in inverse scattering is to reconstruct properties of the hidden objects efficiently and accurately by simultaneously borrowing the strength of different types of solvers. 

To infer the properties of a hidden object from the measurements of scattered waves, forward information is first learned from mathematical models, also known as forward solvers, which  typically involve partial differential equations that describe the propagation of electromagnetic or acoustic waves through the hidden object \citep{kaipio2006statistical}.
There are various types of forward solvers. Among them, nonlinear forward solvers involve nonlinear partial differential equations which are solved by numerical methods such as finite-element methods \citep{CCHbook}. They are relatively accurate but often require significant computational resources and therefore limits their applicability.
%To infer the properties of a hidden object from the measurements of scattered waves, forward solvers are often used as a key component in the solution of inverse scattering problems, which are mathematical algorithms or computational techniques that simulate the propagation of electromagnetic or acoustic waves through the hidden object \citep{kaipio2006statistical}. 
%Among the forward solvers, nonlinear forward solvers are essential tools in many practical applications for understanding and predicting the behavior of waves in nonlinear objects, which are often implemented using numerical methods, such as  finite-element methods \citep{CCHbook}. Nonlinear forward solvers are known for their high accuracy, but they often require significant computational resources, which limits their practicality. 
Alternatively, a commonly used approach is to approximate nonlinear equations by a linearized version of the problems, such as the Born approximation \citep{kazei2018, muhumuza2018}. While these approximations are computationally efficient, they are less accurate than nonlinear solvers. The goal in inverse scattering is to efficiently and accurately reconstruct the properties of a hidden object by leveraging the strengths of various types of solvers simultaneously.
   
%There are extensive studies on inverse scattering problems in applied mathematics and statistics \citep{colton2006using,kaipio2006statistical,cakoni2014qualitative,kaipio2019}. Recent advances in computing have created new opportunities to study inverse problems by allowing more complex mathematical models involving functional inputs to beperformed. This has led to increased attention towards inverse problems with functional inputs. inverse problems with functional inputs have received increasing attention. Take an inverse scattering problem in Figure \ref{fig:inversescattering} as the example where the objective is, given measured far-field pattern, to determine the inhomogeneous material properties as a function on an unknown scatterer, which are of great interest in science, engineering, and medicine. Let the functional input $g$ represent the  material properties of an inhomogeneous scattering region of interest shown in the middle of Figure \ref{fig:inversescattering}. For a given functional input, the far-field pattern, $u^s$, can be simulated through a computer model consisting of systems of partial differential equations \citep{CCHbook}, which can be solved by numerical methods, such as the finite element methods (FEMs).  Based on the training data from the computer simulations, the final goal is to reconstruct and infer the functional input $g$ based on some observed far-field measurements.

\begin{figure}[h]
    \centering
    \includegraphics[width=0.5\textwidth]{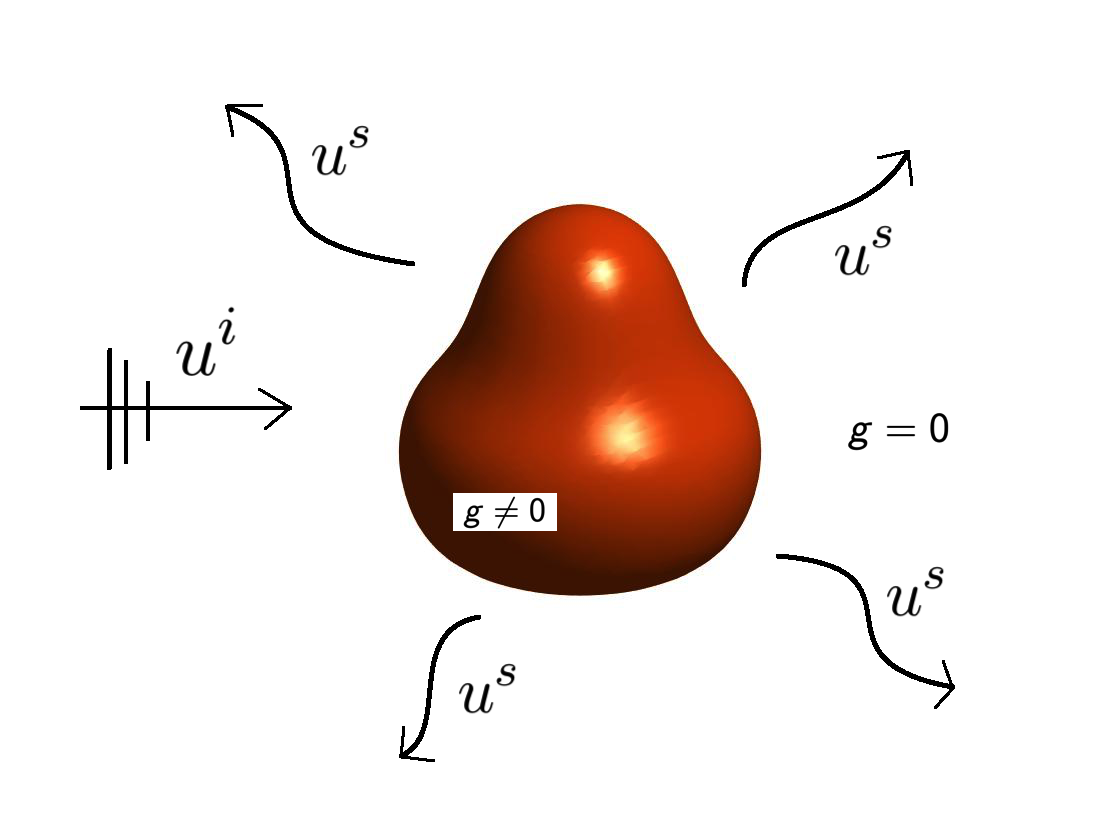}
    \caption{Inverse scattering problems.}
    \label{fig:inversescattering}
\end{figure}

There are extensive studies on inverse scattering problems in applied mathematics and statistics \citep{colton2006using,kaipio2006statistical,cakoni2014qualitative,kaipio2019}. Recent advances in computing have created new opportunities to study inverse problems by allowing more complex mathematical models involving \textit{functional inputs} to be
performed. This has led to an increased attention towards inverse problems with functional inputs which reveal more detailed information regarding the hidden object. For instance, consider an inverse scattering problem illustrated in Figure \ref{fig:inversescattering},  where the objective is to recover the inhomogeneous material properties (denoted by $g$) in a functional form for the unknown scatterer in the middle of Figure \ref{fig:inversescattering}, given the response far-field pattern $u^s$. For any functional input $g$, the far-field pattern, $u^s$, can be simulated through a computer model consisting of systems of partial differential equations \citep{CCHbook}, which can be solved by numerical methods, such as the finite element methods.  Given the training data from computer simulations, the final goal is to reconstruct and infer the functional input $g$ based on  observed far-field measurements.

Despite numerous studies on inverse problems, most of the existing results in the literature are not applicable for making inferences about unknown functional inputs, including the estimation and its uncertainty quantification.
Recent developments for functional inputs have primarily relied on the idea of truncated basis expansion  \citep{tan2019gaussian, li2021gaussian}, which is intuitive and commonly used in functional data analysis but can result in inefficient estimation and additional uncertainty in the inverse estimation. To successfully address the inverse problems with functional inputs, a critical yet challenging step is to construct efficient surrogate models for mathematical forward solvers that can accurately capture the information from  functional inputs and  provide rigorous quantification of the prediction uncertainty.

%Despite numerous studies on inverse problems, most of the existing results are not applicable to make inference for the unknown functional inputs, including the estimation and uncertainty quantification of the estimated functions. Recent developments for functional inputs are mainly based on the idea of basis truncation which are intuitive and commonly used in functional data analysis but often lead to inefficient estimation and additional uncertainty in the inverse estimation \citep{tan2019gaussian, li2021gaussian}. To successfully address the inverse problems with functional inputs, a critical but challenging step is to incorporate surrogate models that can accurately capture the information from functional inputs and therefore provide a rigorous quantification of the estimation uncertainty.

%Motivated by an inverse scattering problem where the functional input of interest is the refractive index over a bounded scatterer, a new Bayesian framework is proposed. This framework tackles the aforementioned challenges by developing a Bayesian procedure incorporating a surrogate model that directly accounts for the functional inputs; therefore, the inverse can be inferred efficiently and accurately by the posterior distribution. 

Motivated by an inverse scattering problem where the functional input of interest is the refractive index over a bounded scatterer, a novel Bayesian framework is proposed. This framework tackles the aforementioned challenges by \textit{constructing an efficient  surrogate model that directly accounts for the functional inputs and a Bayesian procedure that allows efficient and accurate inference of the functional inverse through the posterior distribution}. In particular, the surrogate model is constructed by a Gaussian process prior with functional inputs, which captures the information from the functional input directly through kernel functions without finite basis expansion, % commonly used in functional data analysis \citep{marzouk2009dimensionality,li2021gaussian}, 
thereby retaining the information without any truncation.  Furthermore, this surrogate model is used to efficiently integrate multi-fidelity simulations, including low-fidelity simulations (such as Born approximation) and high-fidelity simulations (such as nonlinear equations solved by finite element methods), to achieve better reconstruction accuracy and computational efficiency. While there have been numerous developments on multi-fidelity emulation \citep{kennedy2000predicting}, most of the existing work focuses on scalar inputs and the extensions to functional inputs are nontrivial. % and developments on functional inputs are scarce. %To the best of our knowledge, this is the first statistical model that attempts to extend the work of \cite{kennedy2000predicting} to multi-fidelity emulation with functional inputs.

Note that, identifying the functional inverse in the current setting is different from the calibration of computer models with functional parameters in recent studies \citep{plumlee2016calibrating,brown2018nonparametric,tuo2021reproducing,sung2022estimating}. The calibration parameters of interest in those studies were represented as functions of control variables, and each model output was generated through a scalar parameter assumed to be a realization of an unknown function $g(x)$.  In contrast, the inverse scattering problem herein involves a single function $g$, which is explicitly given (e.g., $g(x)=1+x$), that generates \textit{only one} response pattern in the computer model. Consequently, there is significantly less information available to guide the search for the inverse in the functional space, leading to a more challenging problem.

The remainder of the paper is organized as follows. 
Section 2 introduces the Bayesian framework for the inverse scattering problem when only one numerical simulator is available. In Section 3, we propose a method to reconstruct the inverse function by integrating multi-fidelity simulators. Section 4 presents the results of inverse prediction and uncertainty quantification for the refractive index of a bounded scatterer. Finally, in Section 5, we conclude with future research directions. Detailed algorithms are provided in Appendix, and the \texttt{R} \citep{R2018} code for reproducing numerical results is provided in Supplemental Materials.

\section{Bayesian inference for functional inverse}

%The goal of inverse scattering is to efficiently and accurately recover/reconstruct the functional input from an observed far-field pattern and provide uncertainty quantification for the reconstructed inputs. 

\begin{figure}[h]
    \centering
    \includegraphics[width=\textwidth]{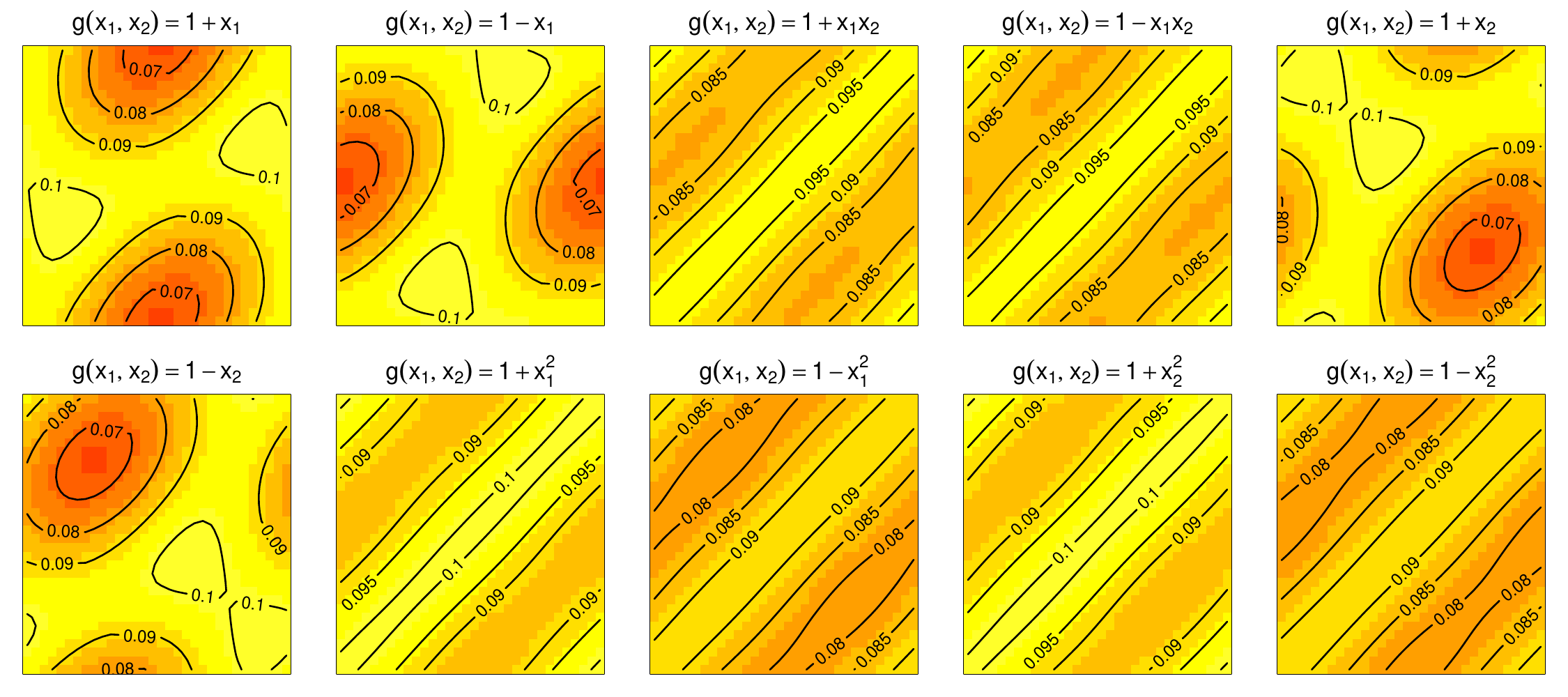}
    \caption{Training data from finite-element simulations, where the functional inputs are given in the panel titles and the images are their corresponding far-field outputs.}
    \label{fig:realcase_real}
\end{figure}

Suppose that $V$ is a functional input space consisting of functions defined on a compact and convex region $\Omega\subseteq\mathbb{R}^d$, and all functions $g(\mathbf{x})\in V$ are continuous in $\Omega$, where $\mathbf{x}\in \Omega$. %i.e., $V\subset C(\Omega)$. 
Given a functional input $g(\mathbf{x})\in V$, $\mathbf{y}^s(g)$ is the corresponding output obtained from a forward mathematical solver which is often computationally intensive, such as a series of nonlinear partial differential equations solved by finite element methods (FEMs). For each functional input, the output is typically an image and it can be vectorized as a vector with length $m$, thus we have $\mathbf{y}^s:V\rightarrow\mathbb{R}^{m}$. For example, based on the ten training input functions shown in the panel titles of Figure \ref{fig:realcase_real}, the corresponding ten outputs $\mathbf{y}^s(g)$ are given in image format as in Figure \ref{fig:realcase_real}, which can be viewed as a vector of length $m=1024$.  Let $\mathbf{y}^p$ be a vector of the observed far-field pattern. It is assumed that    
\begin{equation}\label{eq:inversemodel}
    \mathbf{y}^{p}=\mathbf{y}^s(g)+\mathbf{e},
\end{equation} 
where $\mathbf{e}$ is the random noise and each of its element follows an identical, independent normal distribution with zero mean and variance $\sigma^2_e$, i.e.,
$
    \mathbf{e}\sim\mathcal{N}_{m}(\mathbf{0}_m,\sigma^2_e \mathbf{I}_{m})
$
and $\mathbf{I}_{m}$ is an identity matrix of size $m$. 
Based on a training set received from mathematical forward solvers, the objective in inverse scattering is to reconstruct the functional input $g$ corresponding to the received far-field pattern. 

To develop a Bayesian approach for inverse scattering problems, a surrogate model for mathematical solvers is first introduced in Section 2.1 to efficiently emulate $\mathbf{y}^s(g)$ with functional inputs. This step is crucial because mathematical solvers are typically computationally intensive; therefore, it is infeasible to explore the entire functional input space by mathematical solvers. 
Based on \eqref{eq:inversemodel} and the surrogate model constructed in Section 2.1, a Bayesian framework for inverse scattering is proposed in Section 2.2.

\subsection{Surrogate model with functional inputs  % for $\mathbf{y}^s(g)$
}\label{sec:FEMsurrogate}

Since the response far-field patterns $\mathbf{y}^s(g)$ are images, we first pre-process the images by a dimension reduction method. Specifically, principal component analysis (PCA) is performed to identify the first $L$ principal components, $\mathbf{u}_l\in\mathbb{R}^{1024},l=1,2,\ldots,L$, that explain more than $99\%$ of the variability of the images in the training set.  
%To this end, the pre-processing of using principal component analysis (PCA) is first applied to reduce the dimension of the output images of $\mathbf{y}^s(g)$. The first $L$ principal components, denoted by $\mathbf{u}_l\in\mathbb{R}^{1024\times 1},l=1,2,\ldots,L$, are obtained which explain more than, say 99\%, variability of the data.
As a result, given the functional input $g_i$ for $i=1,\ldots, n$, the output of the far-field images can be  approximated by 
\begin{equation}\label{PCscore}
\mathbf{y}^s(g_i)\approx\sum^L_{l=1}f_l(g_i)\mathbf{u}_l,
\end{equation} 
where $\{f_l(g_i)\}^L_{l=1}$ are the first $L$ principal component scores, which are computed by $f_l(g_i)=\mathbf{u}^T_l\mathbf{y}^s(g_i)$. After the dimension reduction, instead of modeling the output images, the objective becomes to construct an efficient surrogate model that predicts the principal component scores $f_l(g)$ for any $g\in V$.

 Gaussian process (GP) models \citep{santner2003design,gramacy2020surrogates} are widely used to construct statistical surrogates for mathematical solvers, but building a surrogate for $f_l(g)$ is challenging when the input $g$ is in a functional form instead of a scalar input. This is because a new kernel function, denoted by $K_l(g,g')$ with $g,g'\in V$, is needed to describe the correlation structure in a functional space directly. To this end, we adopt the functional-input GP model proposed by \cite{sung2022functional}:
$$
f_l(g)\sim\mathcal{FIGP}(\mu_l,K_l(g,g')),
$$
 where $\mu_l$ is an unknown mean, $K_l(g,g')$ is semi-positive definite, and $\{f_l(g)\}^L_{l=1}$ are assumed to be mutually independent. As introduced by \cite{sung2022functional}, there are two classes of kernel functions for any functional inputs $g_1,g_2\in V$, including a  \textit{linear kernel} :
\begin{align}\label{eq:lk}
    K_l(g_1,g_2)=\tau^2_l\int_{\Omega}\int_{\Omega} g_1(\mathbf{x})g_2(\mathbf{x}')\Phi_{\boldsymbol{\theta}_l}(\mathbf{x},\mathbf{x}'){\rm{d}}\mathbf{x}{\rm{d}}\mathbf{x}',
\end{align}
and a \textit{nonlinear kernel}:
\begin{align}\label{eq:nonlinearkernel}
    K_l(g_1,g_2) = \tau^2_l\phi(\gamma_l\|g_1-g_2\|_{L_2(\Omega)}),
\end{align}
where $\tau_l^2$ is a positive scalar, $\Phi_{\boldsymbol{\theta}_l}$ is a positive definite kernel function defined on $\Omega\times \Omega$ with the hyperparameter $\boldsymbol{\theta}_l$, $\phi(r):\mathbb{R}^{+}\rightarrow\mathbb{R}$ is a radial basis function whose corresponding kernel in $\mathbb{R}^d$ is strictly positive definite for any $d\geq 1$, $\|\cdot\|_{L_2(\Omega)}$ is the $L_2$-norm of a function, defined by $\|h\|_{L_2(\Omega)}=(\langle h,h\rangle_{L_2(\Omega)})^{1/2}$, and $\gamma_l>0$ is a parameter that controls the decay of the kernel function with respect to the $L_2$-norm. Throughout this paper, the kernel function $\Phi_{\boldsymbol{\theta}_l}$ is assumed to be a Mat\'ern kernel, which is widely used in the literature \citep{santner2003design,stein2012interpolation}. The Mat\'ern kernel function has the form of
\begin{eqnarray}\label{eq:matern1}
\Phi_{\boldsymbol{\theta}_l}(\mathbf{x},\mathbf{x}')=\phi(\|\boldsymbol{\theta}_l\odot(\mathbf{x}-\mathbf{x}')\|_2)
\end{eqnarray}
with the Mat\'ern radial basis function:
\begin{eqnarray}\label{eq:matern2}
\phi(r)=
\frac{1}{\Gamma(\nu)2^{\nu-1}}(2\sqrt{\nu} r)^\nu B_\nu(2\sqrt{\nu}r),
\end{eqnarray}
where $\odot$ is the Hadamard product,  and $\boldsymbol{\theta}_l$ is a lengthscale parameter of length $d$, $\|\cdot\|_2$ denotes the Euclidean norm, $B_\nu$ is the modified Bessel function of the second kind, and $\nu$ represents a smoothness parameter. Quasi-Monte Carlo integration \citep{morokoff1995quasi} can be used to numerically evaluate the integrals in the kernels. Specifically, suppose that $\Omega$ is a unit cube, then the linear kernel \eqref{eq:lk} can be approximated by
\begin{equation}\label{eq:linearkernelMC}
K_l(g_1,g_2)\approx\frac{\tau^2_l}{N^2} \left(\mathbf{g}^T_{1,N}\boldsymbol{\Phi}_{\boldsymbol{\theta}_l}\mathbf{g}_{2,N}\right),
\end{equation}
where $\boldsymbol{\Phi}_{\boldsymbol{\theta}_l}$ is an $N\times N$ matrix with each element $(\boldsymbol{\Phi}_{\boldsymbol{\theta}_l})_{i,j}=\Phi_{\boldsymbol{\theta}_l}(\mathbf{x}_i,\mathbf{x}_j)$, $\mathbf{g}_{j,N}$ is a vector of length $N$, which is $\mathbf{g}_{j,N}=(g_j(\mathbf{x}_1),\ldots,g_j(\mathbf{x}_N))^T$, and $\{\mathbf{x}_i\}^N_{i=1}$ is a low-discrepancy sequence from a unit cube, for which the Sobol sequence \citep{sobol1967distribution,bratley1988algorithm} is adopted here.

Based on a selected  kernel function, the $n$ outputs ($f_l(g_1),\ldots,f_l(g_n)$) follow a multivariate normal distribution, 
$$\mathbf{f}_l:=(f_l(g_1),\ldots,f_l(g_n))^T\sim\mathcal{N}_n(\mu_l\boldsymbol{1}_n,\mathbf{K}_l),$$ 
where the covariance $\mathbf{K}_l\in\mathbb{R}^{n\times n}$ with $(\mathbf{K}_l)_{j,k}=K_l(g_j,g_k)$, and $\boldsymbol{1}_n$ is a size-$n$ all-ones vector. 
Denote $\mathbf{Y}_s=(\mathbf{y}^s(g_1),\ldots,\mathbf{y}^s(g_n))\in\mathbb{R}^{m\times n}$, which implies that $\mathbf{f}_l=\mathbf{Y}^T_s\mathbf{u}_l$. The unknown parameters, including $\mu_l$ and the hyperparameters associated with the kernel function, which are either $\boldsymbol{\theta}_l$ in \eqref{eq:lk} or $\gamma_l$ in \eqref{eq:nonlinearkernel}, can be estimated by maximizing the log-likelihood function:
$$
\text{constant}-\frac{1}{2}\log|\mathbf{K}_l|-\frac{1}{2}(\mathbf{Y}^T_s\mathbf{u}_l-\mu_l\boldsymbol{1}_n)^T\mathbf{K}_l^{-1}(\mathbf{Y}^T_s\mathbf{u}_l-\mu_l\boldsymbol{1}_n).
$$

The choice of the kernel function depends on the complexity of the underlying structure. To find the kernel function that balances the bias--variance trade-off in practice, the idea of leave-one-out cross-validation (LOOCV) suggested by \cite{sung2022functional} is implemented, which provides an efficient closed-form expression to select the kernel by minimizing the estimated prediction error:
\begin{equation}\label{eq:LOOCV}
\text{LOOCV}=\frac{1}{n}\| \boldsymbol{\Lambda}_l^{-1}\mathbf{K}_l^{-1}(\mathbf{Y}^T_s\mathbf{u}_l-\mu_l\boldsymbol{1}_n)\|^2_2,
\end{equation}
where 
$\boldsymbol{\Lambda}_l\in\mathbb{R}^{n\times n}$ is a diagonal matrix with the element $(\boldsymbol{\Lambda}_l)_{j,j}=(\mathbf{K}_l^{-1})_{j,j}$.

By the property of the conditional multivariate normal distribution, the corresponding output $f_l(g)$ for an untried functional input, $g\in V$, follows a normal distribution
\begin{equation}\label{fl}
f_l(g)|\mathbf{Y}_s\sim\mathcal{N}(m_l(g), v_l(g))
\end{equation}
with 
\begin{equation}\label{eq:mean}
    m_l(g)= \mu_l+\mathbf{k}_l(g)^T \mathbf{K}^{-1}_l (\mathbf{Y}^T_s\mathbf{u}_l-\mu_l\boldsymbol{1}_n)
\end{equation}
and
\begin{equation}\label{eq:var}
v_l(g)=  K_l(g,g)-\mathbf{k}_l(g)^T \mathbf{K}^{-1}_l \mathbf{k}_l(g),
\end{equation}
where $\mathbf{k}_l(g)=(K_l(g,g_1),...,K_l(g,g_n))^T\in\mathbb{R}^{n\times 1}$. Hence, since $\mathbf{y}^s(g_i)\approx\sum^L_{l=1}f_l(g_i)\mathbf{u}_l$,  the surrogate model $\mathbf{y}^s(g)$ follows a multivariate normal distribution as 
\begin{equation}\label{eq:ysdist}
\mathbf{y}^s(g)\sim\mathcal{N}_m\left(\sum^L_{l=1}m_l(g)\mathbf{u}_l, \sum^L_{l=1}v_l(g)\mathbf{u}_l\mathbf{u}^T_l\right).
\end{equation}

\subsection{Bayesian approach for functional inverse}\label{sec:bayesianinverse}

Given the output $\mathbf{y}^p$ in the model \eqref{eq:inversemodel}, we assume that the unknown functional inverse follows a GP prior. By combining \eqref{eq:inversemodel}, \eqref{eq:ysdist}, and the prior information, the following  Bayesian framework is considered,
\begin{align}
    \mathbf{y}^{p}|g,\sigma^2_e&\sim \mathcal{N}_m\left(\sum^L_{l=1}m_l(g)\mathbf{u}_l, \sigma^2_e \mathbf{I}_{m}+\sum^L_{l=1}v_l(g)\mathbf{u}_l\mathbf{u}^T_l\right),\label{eq:likelihoodmodel}\\
    g(\mathbf{x})|\boldsymbol{\eta},\sigma^2_g&\sim \mathcal{GP}(0,\tau^2_g\Phi_{\boldsymbol{\eta}}(\mathbf{x},\mathbf{x}')),\label{eq:GPprior}\\
    \sigma^2_e&\sim \text{InvGamma}(a_1,b_1),\label{eq:s2prior}\\
    \tau^2_g&\propto \text{InvGamma}(a_2,b_2),\label{eq:taugprior}\\
    \eta_j &\propto \text{Gamma}(a_3,b_3) \quad\text{for}\quad j=1,\ldots,d,\label{eq:etaprior}
\end{align}
where $\text{InvGamma}(a,b)$ denotes an inverse gamma distribution with shape parameter $a$ and rate parameter $b$. In \eqref{eq:GPprior}, we assume that the functional input $g$ follows a GP prior, denoted by $\mathcal{GP}$, where $\Phi_{\boldsymbol{\eta}}$ is a Mat\'ern kernel as in  \eqref{eq:matern1} with the lengthscale parameter $\boldsymbol{\eta}$. The prior distributions of the parameters $\sigma_e^2$, $\tau_g^2$, and $\boldsymbol{\eta}$ are given in \eqref{eq:s2prior}, \eqref{eq:taugprior}, and \eqref{eq:etaprior}, respectively.

%The idea of placing a GP prior on a functional parameter is also adopted in the calibration problems in the computer experiment literature \citep{plumlee2016calibrating,brown2018nonparametric,sung2022estimating}. 

Note that the GP prior is also used to model parameters or inputs that are represented as a function in the literature of inverse problems; however, most of existing works 
%consider a GP with a \textit{fixed} lengthscale parameter $\eta_j$ \citep{plumlee2016calibrating,kaipio2019} or 
consider a truncated Karhunen--Lo\`eve (KL) expansion to reduce the dimension of the GP prior  \citep{marzouk2009dimensionality,li2021gaussian,yang2017bayesian}. In contrast, the proposed model employs a GP prior that includes a prior density for the lengthscale parameter as in \eqref{eq:etaprior} and avoids basis expansion and truncation, which is important because finite truncation associated with basis expansion can introduce model bias. 
In addition, based on the proposed emulator \eqref{eq:ysdist}, the functional-input emulator is modeled directly through kernel functions without finite basis expansion. Therefore, the input information is better preserved  compared to \cite{tan2019gaussian} and \cite{li2021gaussian}, where the emulator is built based on a finite basis expansion of the functional inputs. 
%It is possible to achieve this because the emulator developed as in \eqref{eq:ysdist}  \textit{directly} handles the functional inputs without requiring a finite basis expansion, which sets it apart from \cite{tan2019gaussian} and \cite{li2021gaussian} where the emulator is built based upon finite basis expansion on the functional inputs. 
Numerical comparisons are available in Section \ref{sec:application}. %indicating that utilizing the basis-expansion approach can lead to model bias, resulting in inferior performance.

The computation involved in the likelihood function of \eqref{eq:likelihoodmodel} can further be simplified as follows. By the Woodbury matrix identity \citep{harville1998matrix} and the fact that $\mathbf{u}^T_l\mathbf{u}_l=1$ and $\mathbf{u}^T_l\mathbf{u}_{l'}=0$ for any $l\neq l'$, the covariance inverse of \eqref{eq:likelihoodmodel} can be written as
$$
\left(\sigma^2_e \mathbf{I}_{m}+\sum^L_{l=1}v_l(g)\mathbf{u}_l\mathbf{u}^T_l\right)^{-1}=\frac{1}{\sigma^2_e}\mathbf{I}_{m}-\frac{1}{\sigma^2_e}\mathbf{U}\left(\sigma^2_e\mathbf{V}_L(g)^{-1}+\mathbf{I}_{L}\right)^{-1} \mathbf{U}^T
$$
where $\mathbf{V}_L(g)\in\mathbb{R}^{L\times L}$ is a diagonal matrix with each element $(\mathbf{V}_L(g))_l=v_l(g)$, and the determinant of the covariance can be simplified as 
$$
\text{det}\left(\sigma^2_e \mathbf{I}_{m}+\sum^L_{l=1}v_l(g)\mathbf{u}_l\mathbf{u}^T_l\right)=(\sigma_e^2)^m\prod^L_{l=1}\left(1+\frac{v_l(g)}{\sigma_e^2}\right).
$$
As a result, the likelihood function of \eqref{eq:likelihoodmodel}  can be written as
\begin{align}\label{eq:likelihood}
    L(\mathbf{y}^{p}|g,\sigma^2_e)\propto&(\sigma_e^2)^{-\frac{m}{2}}\prod^L_{l=1}\left(1+\frac{v_l(g)}{\sigma_e^2}\right)^{-1/2}\times\nonumber\\
    &\exp\left\{-\frac{1}{2\sigma^2_e} \left(\left\|\mathbf{y}^{p}-\sum^L_{l=1}m_l(g)\mathbf{u}_l\right\|^2_2-\sum^L_{l=1}\frac{(\mathbf{u}^T_l\mathbf{y}^{p}-m_l(g))^2}{\sigma^2_e/v_l(g)+1}\right)\right\}.
\end{align}
The expression in \eqref{eq:likelihood} allows the likelihood function of \eqref{eq:likelihoodmodel} to be computed easily because it circumvents the need of the $m\times m$ matrix inversion associated with the $m$-dimensional multivariate normal distribution.

%Suppose that $\mathbf{X}_N=\{\mathbf{x}_i\}^N_{i=1}$ is the low-discrepancy sequence used to approximate the kernel function as in \eqref{eq:linearkernelMC}. 

To evaluate the likelihood function $L(\mathbf{y}_n^p|g,\sigma^2_e)$, the function $g$ is replaced by its realization $\mathbf{g}_N=(g(\mathbf{x}_1), \ldots,g(\mathbf{x}_N))^T$ at a low-discrepancy sequence $\mathbf{X}_N=\{\mathbf{x}_i\}^N_{i=1}$, which enables the evaluations of $m_l(g)$ and $v_l(g)$ in the likelihood function, in which the kernel function $\mathbf{k}_l(g)$ of \eqref{eq:mean} and \eqref{eq:var} is approximated based on $\mathbf{g}_N$ as in \eqref{eq:linearkernelMC}. Thus, given the observation $\mathbf{y}^p$, the posterior of the functional input $g(\mathbf{x})$ can  be obtained by 
\begin{align}\label{postg}
&\pi(g(\mathbf{x}),\mathbf{g}_N,\sigma^2_e,\boldsymbol{\eta},\tau^2_g|\mathbf{y}^p)\propto \pi(g(\mathbf{x})|\mathbf{g}_N,\boldsymbol{\eta},\tau^2_g)\pi(\mathbf{g}_N,\sigma^2_e,\boldsymbol{\eta},\tau^2_g|\mathbf{y}^p).
\end{align}
The joint posterior distribution of $g(\mathbf{x})$ can be approximated by Markov chain Monte Carlo (MCMC) by drawing the samples from $\pi(g(\mathbf{x})|\mathbf{g}_N,\boldsymbol{\eta},\tau^2_g)$ and $\pi(\mathbf{g}_N,\sigma^2_e,\boldsymbol{\eta},\tau^2_g|\mathbf{Y}_n^p)$, iteratively. For the posterior $\pi(\mathbf{g}_N,\sigma^2_e,\boldsymbol{\eta},\sigma^2_g|\mathbf{y}^p)$, it follows that 
\begin{align}\label{eq:jointdist}
&\pi(\mathbf{g}_N,\sigma^2_e,\boldsymbol{\eta},\tau^2_g|\mathbf{y}^p)\propto L(\mathbf{y}_n^p|\mathbf{g}_N,\sigma^2_e)\pi(\mathbf{g}_N|\boldsymbol{\eta},\tau^2_g)\pi(\sigma^2_e)\pi(\boldsymbol{\eta})\pi(\tau^2_g)
\end{align}
where $L(\mathbf{y}_n^p|\mathbf{g}_N,\sigma^2_e)$ is the likelihood function \eqref{eq:likelihood}  with the realization $\mathbf{g}_N$,  $\pi(\mathbf{g}_N|\boldsymbol{\eta},\tau^2_g)$ follows $\mathcal{N}_N(\mathbf{0}_N,\tau^2_g\boldsymbol{\Phi}_{\boldsymbol{\eta}})$ based on the GP prior \eqref{eq:GPprior}, where $\boldsymbol{\Phi}_{\boldsymbol{\eta}}$ is an $N\times N$ matrix with each element $(\boldsymbol{\Phi}_{\boldsymbol{\eta}})_{i,j}=\Phi_{\boldsymbol{\eta}}(\mathbf{x}_i,\mathbf{x}_j)$, and $\pi(\sigma^2_e)$, $\pi(\boldsymbol{\eta})$, and $\pi(\tau^2_g)$ are the priors as in \eqref{eq:s2prior}, \eqref{eq:taugprior}, and \eqref{eq:etaprior}. The samples from this posterior distribution can be drawn by Gibbs sampling with Metropolis-Hastings algorithm. The details are given in Appendix \ref{sec:suppsampler}.

The posterior $\pi(g(\mathbf{x})|\mathbf{g}_N,\boldsymbol{\eta},\tau^2_g)$ can be drawn based on the property of conditional multivariate normal distributions, that is,
$$
g(\mathbf{x})|\mathbf{g}_N,\boldsymbol{\eta},\tau^2_g\sim \mathcal{N}(\Phi_{\boldsymbol{\eta}}(\mathbf{x}, \mathbf{X}_N)\boldsymbol{\Phi}_{\boldsymbol{\eta}}^{-1}\mathbf{g}_N, \tau^2_g(1- \Phi_{\boldsymbol{\eta}}(\mathbf{x}, \mathbf{X}_N) \boldsymbol{\Phi}_{\boldsymbol{\eta}}^{-1}\Phi_{\boldsymbol{\eta}}(\mathbf{X}_N,\mathbf{x}))),
$$
where $\Phi_{\boldsymbol{\eta}}(\mathbf{x}, \mathbf{X}_N)=(\Phi_{\boldsymbol{\eta}}(\mathbf{x},\mathbf{x}_1),\ldots,\Phi_{\boldsymbol{\eta}}(\mathbf{x},\mathbf{x}_N))$, and $\Phi_{\boldsymbol{\eta}}(\mathbf{X}_N,\mathbf{x})=\Phi_{\boldsymbol{\eta}}(\mathbf{x}, \mathbf{X}_N)^T$.

%Denote the $N$ realizations of $g(\mathbf{x})$  as $\mathbf{g}_N:=(g(\mathbf{x}_1),\ldots,g(\mathbf{x}_N))$, which follow a multivaraite normal distribution from the GP assumption in \eqref{eq:GPprior}:
%\begin{equation}\label{eq:GPlikelihood}
%\mathbf{g}_N|\boldsymbol{\eta},\tau^2_g\sim\mathcal{%N}_N(\mathbf{0}_N,\tau^2_g\Phi_{\boldsymbol{\eta}}(\mathbf{X}_N,\mathbf{X}_N)),
%\end{equation}
%where $\mathbf{X}_N=\{\mathbf{x}_i\}^N_{i=1}$ is the low-discrepancy sequence used to approximate the kernel function in \eqref{eq:linearkernelMC}.

%$\pi(\mathbf{g}_N|\boldsymbol{\eta},\tau^2_g)$ is the multivariate normal distribution as in \eqref{eq:GPlikelihood}, and $\pi(\sigma^2_e)$, $\pi(\boldsymbol{\eta})$, and $\pi(\tau^2_g)$ are the priors in \eqref{eq:s2prior}, \eqref{eq:taugprior}, and \eqref{eq:etaprior}. 

Similarly, the posterior of $\mathbf{y}^s(g)$ given $\mathbf{y}^p$ and $\mathbf{Y}_s$ can be  obtained by 
\begin{align}\label{postys}
&\pi(\mathbf{y}^s(g),\mathbf{g}_N,\sigma^2_e,\boldsymbol{\eta},\tau^2_g|\mathbf{y}^p,\mathbf{Y}_s)\propto \pi(\mathbf{y}^s(g)|\mathbf{g}_N,\mathbf{Y}_s)\pi(\mathbf{g}_N,\sigma^2_e,\boldsymbol{\eta},\tau^2_g|\mathbf{y}^p),
\end{align}
where the samples from $\pi(\mathbf{g}_N,\sigma^2_e,\boldsymbol{\eta},\tau^2_g|\mathbf{y}^p)$ can be drawn as in \eqref{eq:jointdist}, and the samples from $\pi(\mathbf{y}^s(g)|\mathbf{g}_N,\mathbf{Y}_s)$ can be drawn the multivariate normal distribution \eqref{eq:ysdist} with the realization $\mathbf{g}_N$ for evaluating  the kernel functions.

\section{Surrogate model for multi-fidelity solvers with functional inputs}\label{sec:multisurrogate}

\begin{figure}[h]
    \centering
    \includegraphics[width=\textwidth]{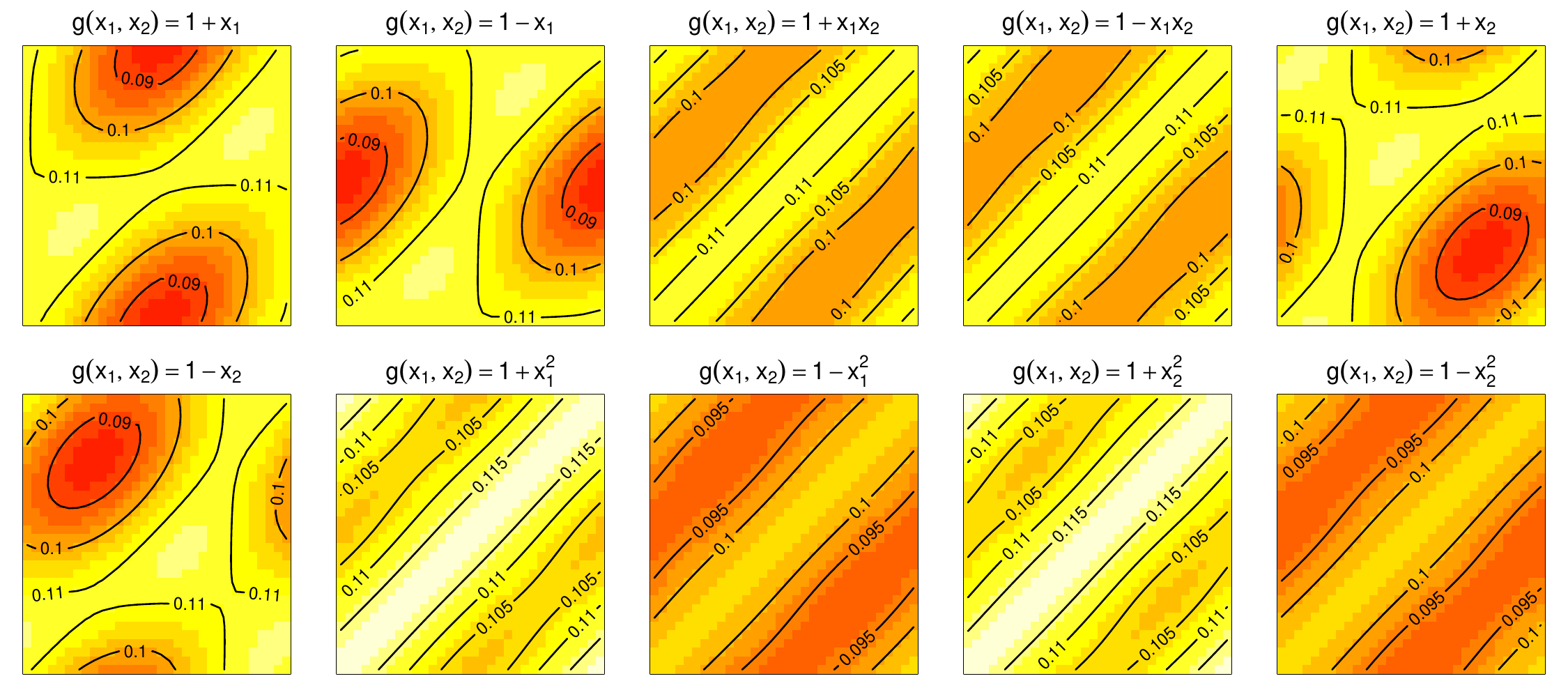}
    \caption{Training data of Born approximation simulator, where the functional inputs are given in the titles and the images are their corresponding far-field outputs.}
    \label{fig:realcase_real_born}
\end{figure}

Section 2 focuses on the situations where only one solver is available for the inverse scattering problem. In practice, there are often multiple solvers available with different  accuracy and different computational efficiency. Accurate forward solvers, also known as high-fidelity simulators, involve nonlinear differential equations that are solved by FEMs, which are often prohibitively costly to have sufficient training samples to explore the functional input space. On the other hand, linearized approximations, such as Born approximation \citep{kazei2018, muhumuza2018}, are low-fidelity solvers, which are developed to serve as faster alternatives, although they are less accurate compared to the high-fidelity simulators. Figure \ref{fig:realcase_real_born} presents the far-field patterns simulated based on Born approximation, which are computationally faster to obtain than the FEM simulations shown in Figure \ref{fig:realcase_real} but less accurate.
A new Bayesian framework, extending from Section 2, is introduced to infer functional inverse in scattering problems by integrating multi-fidelity simulations. We denote the high-fidelity simulator as $\mathbf{y}^s(g)$ and denote the low-fidelity simulator as $\mathbf{y}^b(g)$.   

  %for fitting the emulator model.  %The idea is that, by leveraginguseful information from cheaper lower-fidelity simulations to enhance predictions for the high-fidelity model, an accurate multi-fidelity emulator can be trained with fewer high-fidelity runsand thus lower simulation costs. 
%In this study, besides the FEM simulator $\mathbf{y}^s(g)$,  a lower-fidelity simulator, called \textit{Born approximation} \citep{kazei2018, muhumuza2018}, denoted by $\mathbf{y}^b(g)\in\mathbb{R}^m$, is available as a faster alternative, although less accurate. % for high scattering contrasts. 

%Despite extensive studies on multi-fidelity computer experiments in the literature (see, e.g., \cite{kennedy2000predicting}), the developments are mainly based on scalar inputs instead of functional inputs. In this project, we aim to develop a new framework, including statistical modeling and experimental designs, for multi-fidelity computer experiments. 

%Recall that $\mathbf{y}^s(g_i)\approx\sum^L_{l=1}f_l(g_i)\mathbf{u}_l$, where $f_l(g_i)=\mathbf{u}^T_l\mathbf{y}^s(g_i)$. 

Similar to Section 2, we apply PCA to the image output $\mathbf{y}^b(g)$, where the principal component scores are obtained by $h_l(g)=\mathbf{u}^T_l\mathbf{y}^b(g)$ with the principal component $\mathbf{u}_l$  given in \eqref{PCscore} for $l=1,\ldots,L$. To emulate the high-fidelity simulator, we assume an autoregressive model that integrates the multi-fidelity solvers:
%Since the outputs from both simulators are images, we perform PCA for both $\mathbf{y}^s(g)$ and $\mathbf{y}^b(g)$ and integrate the information through their principal component scores denoted by $f_l(g)$ and $h_l(g)$, where $h_l(g)=\mathbf{u}^T_l\mathbf{y}^b(g)$ and $\mathbf{u}^T$ is given by \eqref{PCscore}. High-fidelity simulators are more accurate than than low-fidelity simulators; therefore, an autoregressive model is assumed: 
%To integrate the high-fidelity simulator $\mathbf{y}^s(g)$ and the Born approximation $\mathbf{y}^b(g)$, we propose an autoregressive model for the PC score $f_l(g)$ as follows, 
\begin{equation}\label{eq:autoregressive}
f_l(g)=\rho_l h_l(g)+\delta_l(g),
\end{equation}
where $\rho_l$ is an unknown parameter and $\delta_l(g)$ is the unknown discrepancy between high-fidelity and low-fidelity simulators. The model can be viewed as an extension of the autoregressive model proposed by \cite{kennedy2000predicting} to functional inputs. Similar to the idea in Section 2, $h_l(g)$ and $\delta_l(g)$ are modeled as a functional-input GP  \citep{sung2022functional}, i.e.,
\begin{equation}\label{eq:multiFIGP}
    h_l(g)\sim\mathcal{FIGP}(\mu_{h_l},K_{h_l}(g,g'))\quad\text{and}\quad \delta_l(g)\sim\mathcal{FIGP}(\mu_{\delta_l},K_{\delta_l}(g,g'))
\end{equation}
for $l=1,\ldots,L$, and assume that $\{h_l(g)\}^L_{l=1}$ and $\{\delta_l(g)\}^L_{l=1}$ are mutually independent. The definitions of $\mu_{h_l},\mu_{\delta_l},K_{h_l}$ and $K_{\delta_l}$ are similar to the ones defined in Section \ref{sec:FEMsurrogate}. Thus, it follows that
$$(h_l(g_1),\ldots,h_l(g_n))^T\sim\mathcal{N}_n(\mu_{h_l}\mathbf{1}_n,\mathbf{K}_{h_l})\quad\text{and}\quad (\delta_l(g_1),\ldots,\delta_l(g_n))^T\sim\mathcal{N}_n(\mu_{\delta_l}\mathbf{1}_n,\mathbf{K}_{\delta_l}),$$ 
where $(\mathbf{K}_{h_l})_{j,k}=K_{h_l}(g_j,g_k)$ and $(\mathbf{K}_{\delta_l})_{j,k}=K_{\delta_l}(g_j,g_k)$. Since $\delta_l(g_i)=f_l(g_i)-\rho_l h_l(g_i)=\mathbf{u}^T_l(\mathbf{y}^s(g_i)-
\rho_l\mathbf{y}^b(g_i))$, the parameters, $\rho_l,\mu_{\delta_l}$ and the parameters associated with the kernel $K_{\delta_l}$, can be estimated by maximizing the log-likelihood function, 
$$
\text{constant}-\frac{1}{2}\log|\mathbf{K}_{\delta_l}|-\frac{1}{2}((\mathbf{Y}_s-
\rho_l\mathbf{Y}_b)^T\mathbf{u}_l-\mu_{\delta_l}\boldsymbol{1}_n)^T\mathbf{K}_{\delta_l}^{-1}((\mathbf{Y}_s-
\rho_l\mathbf{Y}_b)^T\mathbf{u}_l-\mu_{\delta_l}\boldsymbol{1}_n),
$$
where  $\mathbf{Y}_b=(\mathbf{y}^b(g_1),\ldots,\mathbf{y}^b(g_n))\in\mathbb{R}^{m\times n}$. The parameter, $\mu_{h_l}$, and the parameters of the kernel $K_{h_l}$, can be estimated by maximizing the log-likelihood function, 
$$
\text{constant}-\frac{1}{2}\log|\mathbf{K}_{h_l}|-\frac{1}{2}(\mathbf{Y}_b^T\mathbf{u}_l-\mu_{h_l}\boldsymbol{1}_n)^T\mathbf{K}_{h_l}^{-1}(\mathbf{Y}_b^T\mathbf{u}_l-\mu_{h_l}\boldsymbol{1}_n).
$$

By the property of the conditional multivariate normal distribution, the corresponding output $f_l(g)$ for an untried input, $g\in V$, follows a normal distribution
\begin{equation}\label{PostMulti}
f_l(g)|\mathbf{Y}_s,\mathbf{Y}_b\sim\mathcal{N}(m^{(b)}_l(g), v^{(b)}_l(g))
\end{equation}
with 
\begin{equation*}
    m^{(b)}_l(g)= (\rho_l\mu_{h_l}+\mu_{\delta_l})+\mathbf{t}_l(g)^T\mathbf{V}_l^{-1}(\mathbf{z}_l-\boldsymbol{\mu}_l)
\end{equation*}
and
\begin{equation*}
v^{(b)}_l(g)=  \rho^2_lK_{h_l}(g,g)+K_{\delta_l}(g,g)-\mathbf{t}_l(g)^T\mathbf{V}_l^{-1}\mathbf{t}_l(g),
\end{equation*}
where 
$$
\mathbf{z}_l=\left(\begin{array}{c}
     \mathbf{Y}^T_b\mathbf{u}_l  \\
    \mathbf{Y}^T_s\mathbf{u}_l
\end{array}\right),\quad \boldsymbol{\mu}_l=\left(\begin{array}{c}
    \mu_{h_l}\mathbf{1}_n \\
    (\rho_l\mu_{h_l}+\mu_{\delta_l})\mathbf{1}_n  
\end{array}\right),
$$
$$
\mathbf{V}_l=\left(\begin{array}{cc}
   \mathbf{K}_{h_l}  & \rho_l\mathbf{K}_{h_l} \\
   \rho_l\mathbf{K}_{h_l}  & \rho^2_l\mathbf{K}_{h_l}+\mathbf{K}_{\delta_l}
\end{array}\right),\quad\mathbf{t}_l(g)=\left(\begin{array}{c}
   \rho_l\mathbf{k}_{h_l}(g) \\ \rho_l^2\mathbf{k}_{h_l}(g)+\mathbf{k}_{\delta_l}(g)
\end{array}\right),
$$
where $\mathbf{k}_{h_l}(g)=(K_{h_l}(g,g_1),...,K_{h_l}(g,g_n))^T$ and $\mathbf{k}_{\delta_l}(g)=(K_{\delta_l}(g,g_1),...,K_{\delta_l}(g,g_n))^T$. 

According to \eqref{PostMulti}, the Bayesian framework developed in Section 2 can be easily extended to integrate multi-fidelity simulators by replacing the mean $m_l(g)$ and variance $v_l(g)$ by $m_l^{(b)}(g)$ and $v_l^{(b)}(g)$ respectively. 

%approach for inverse scattering problems developed in Section \ref{sec:bayesianinverse} can employ the multi-fidelity surrogate model, instead of the single-fidelity surrogate model in Section \ref{sec:FEMsurrogate}, by replacing the mean $m_l(g)$ and variance $v_l(g)$ with $m_l^{(b)}(g)$ and $v_l^{(b)}(g)$, respectively. Both of the results based on the single-fidelity and multi-fidelity surrogate models will be demonstrated in the next section. 

\section{Applications to Inverse Scattering Problems}\label{sec:application}

The inverse scattering problem introduced in Section \ref{sec:intro} is revisited and analyzed by the proposed Bayesian framework. Based on the training set simulated from mathematical models, the objective is to infer the functional input, given an observed far-field pattern. The functional input $g$ in this study represents the refractive index of the bounded medium as shown in the middle of Figure \ref{fig:inversescattering}.
The proposed Bayesian approach is implemented to determine the refractive index $g$ where the estimation uncertainty is measured using its posterior distribution. Two  analysis results are presented in Sections \ref{sec:resultsingle} and \ref{sec:resultmulti}, with one analysis based solely on FEM simulators and the other based on multi-fidelity simulations (FEM and Born approximation).

\begin{figure}[h]
    \centering
    \includegraphics[width=0.7\textwidth]{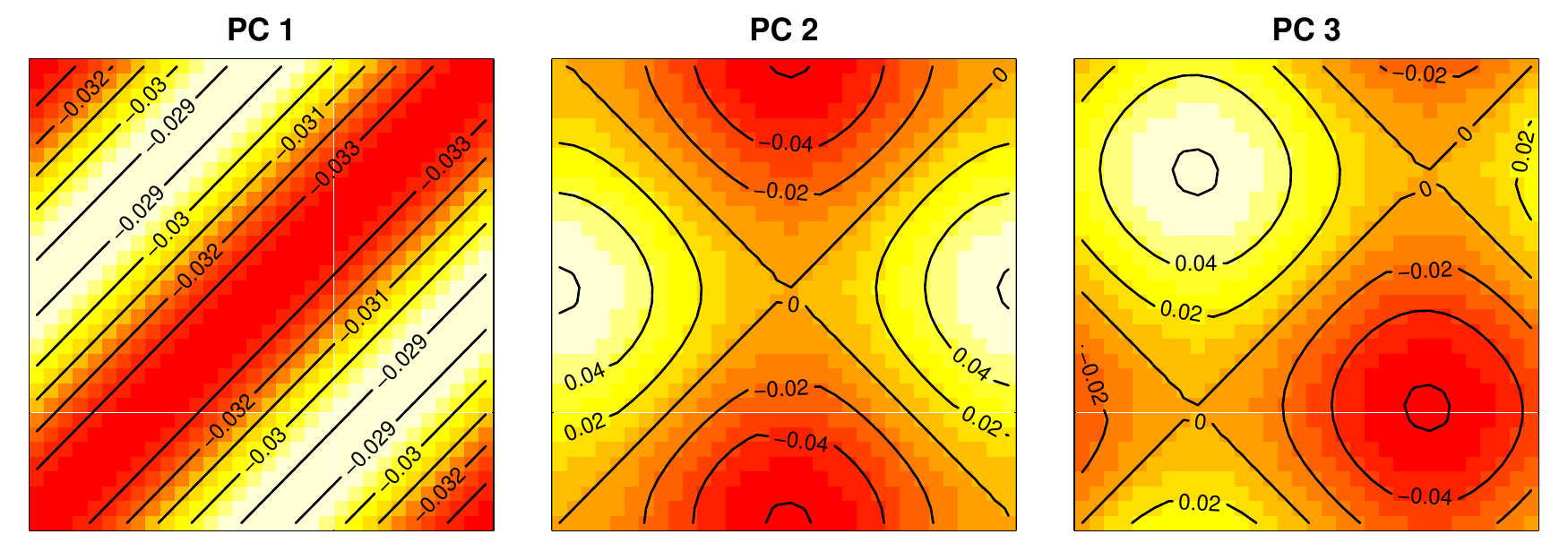}
    \caption{Principle components, which explain more than 99.99\% variations of the data.}
    \label{fig:realcase_pc}
\end{figure}

\begin{figure}[h]
    \centering
    \includegraphics[width=0.7\textwidth]{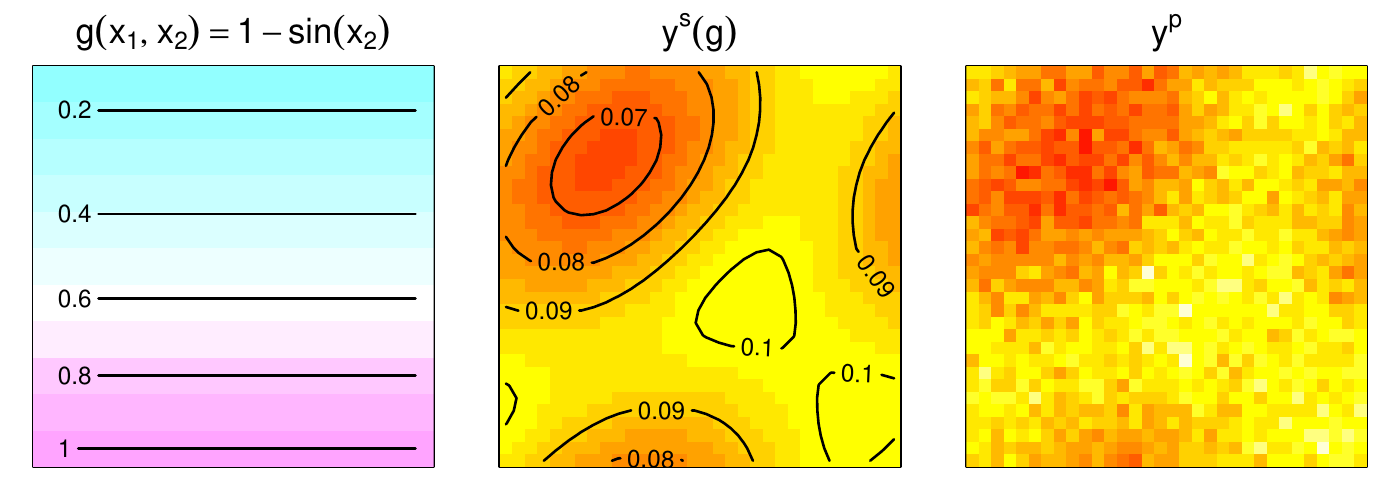}
    \caption{Illustration of the testing data. The underlying functional input (left), the FEM output (middle), and the observed far-field measurement (right).}
    \label{fig:inverse_true}
\end{figure}

In this study, $\Omega\subseteq\mathbb{R}^2$, and 10 functional inputs of $g$, including $1,1+x_1,1-x_1,1+x_1x_2,1-x_1x_2,1+x_2,1+x_1^2,1-x_1^2,1+x_2^2$, and $1-x_2^2$, are considered in the training set and their corresponding FEM simulations are shown in Figure \ref{fig:realcase_real}. 
The PCA is first applied that yields the three principle components, $\mathbf{u}_l\in\mathbb{R}^{1024},l=1,2,3$, explaining more than 99.99\% variability of the data, which are presented in Figure \ref{fig:realcase_pc}. The proposed method is tested on a functional input, $g(x_1,x_2)=1-\sin(x_2)$, which is shown in the left panel of Figure \ref{fig:inverse_true}, and its corresponding output $\mathbf{y}^s(g)$ is presented in the middle panel. The corresponding far-field pattern $\mathbf{y}^p$ is generated according to the model \eqref{eq:inversemodel} with $\sigma^2_e=0.005^2$, which is shown in the right panel of Figure \ref{fig:inverse_true}.

 %Take the functional input $g(x_1,x_2)=1-\sin(x_2)$ as an example, its FEM simulated output $\mathbf{y}^s(g)$  and physical observational output $\mathbf{y}^p$, are illustrated in Figure \ref{fig:inverse_true}.

The hyperparameters in the priors \eqref{eq:s2prior}, \eqref{eq:taugprior}, and \eqref{eq:etaprior} are set as follows: $a_1=a_2=a_3=1$ and $b_1=b_3=1$ and $b_2=10^{-5}$.  The MCMC sampling involves 5000 iterations during a burn-in period, followed by an additional 5000 samples drawn and thinned to minimize autocorrelation. The Mat{\'e}rn kernel function with the smoothness parameter $\nu=5/2$ is considered, which leads to a simplified form of \eqref{eq:matern2}:
\begin{equation}\label{eq:maternkernel2.5}
    \phi(r)=\left(1+\sqrt{5}r+\frac{5}{3}r^2\right)\exp\left(-\sqrt{5}r\right).
\end{equation}
The Sobol sequence \citep{sobol1967distribution,bratley1988algorithm} of size $N=100$ is employed to approximate the kernel functions as in \eqref{eq:linearkernelMC}. 

\subsection{Inverse results with finite-element simulator}\label{sec:resultsingle}

The functional-input GP introduced in Section \ref{sec:FEMsurrogate} is implemented to construct the emulator based on the finite-element simulations $\{\mathbf{y}^s(g_i)\}^{10}_{i=1}$ presented in Figure \ref{fig:realcase_real}. The LOOCV criterion defined in \eqref{eq:LOOCV} are computed for the functional-input GPs, $f_1(g),f_2(g)$ and $f_3(g)$, using the linear kernel \eqref{eq:lk} and nonlinear kernel \eqref{eq:nonlinearkernel}. The results are summarized in Table \ref{tab:loocv_single}, in which the linear kernel is suggested for all the three principal components.

\begin{table}[t]
\begin{center}
\begin{tabular}{ c|c|C{2.6cm}|C{2.6cm}|C{2.6cm} } 
 \toprule
 & Kernel  & $f_1(g)$ & $f_2(g)$& $f_3(g)$ \\ 
 \midrule
 \multirow{2}{*}{LOOCV} &  linear& $\boldsymbol{7.5\times 10^{-5}}$ & $\boldsymbol{6.1\times 10^{-16}}$ & $\boldsymbol{7.6\times 10^{-16}}$\\ 
 & nonlinear& $8.1\times 10^{-5}$ & $2.2\times 10^{-6}$ & $1.9\times 10^{-6}$\\ 
 \bottomrule
\end{tabular}
\end{center}
    \caption{The leave-one-out cross-validation errors (LOOCVs) for each of the functional-input GPs, $f_l(g)$, based on the FEM simulations. The errors corresponding to the optimal kernel are boldfaced.}
    \label{tab:loocv_single}
\end{table}

\begin{figure}[t]
    \centering
    \includegraphics[width=\textwidth]{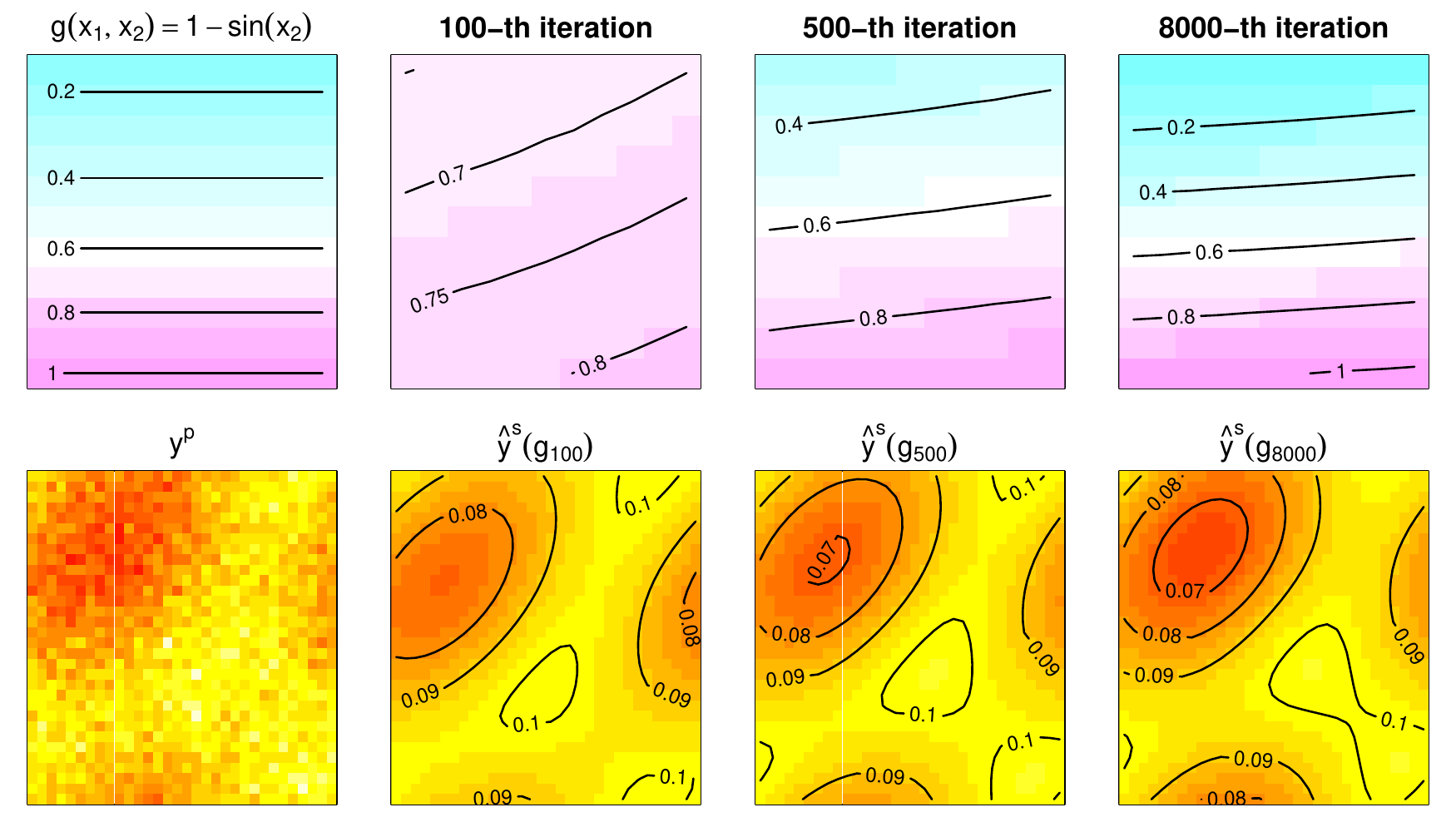}
    \caption{Demonstration of the MCMC algorithm. Upper panels show (for left to right) the true function $g$, and the sampled $g$ in the 100-th, 500-th, and 8000-th iterations of the MCMC algorithm. Lower panels show (for left to right) the physical far-field pattern $\mathbf{y}^p$, and the corresponding prediction means of $\mathbf{y}^s(g)$ with the sampled $g$. }
    \label{fig:inverse_demo}
\end{figure}

\begin{figure}[h]
    \centering
    \includegraphics[width=\textwidth]{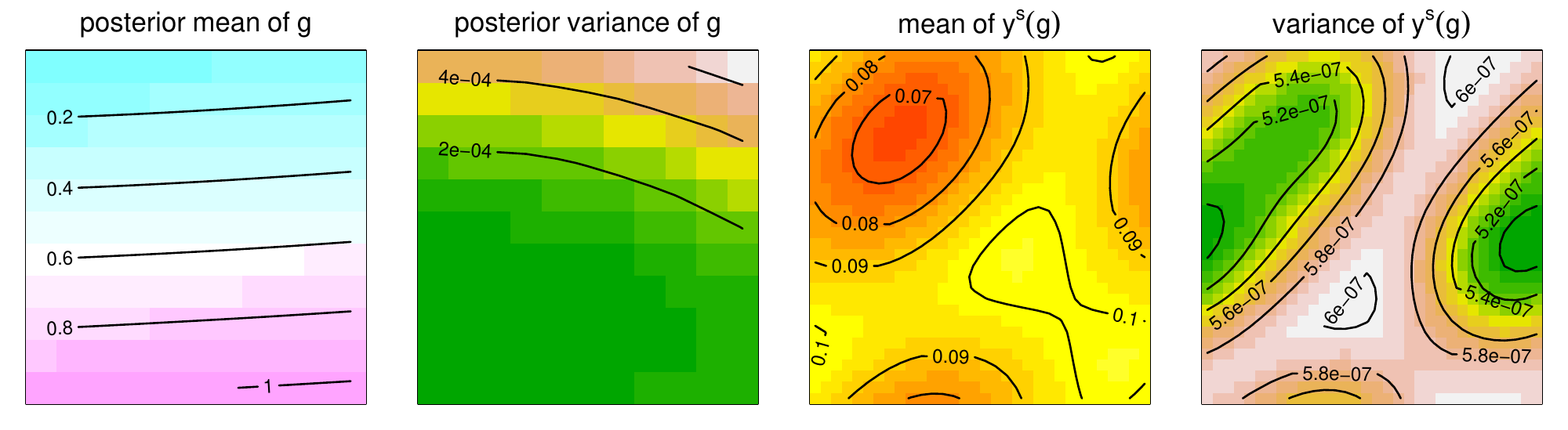}
    \caption{The inverse result based on the FEM simulator. The left two panels show the mean and variance of the MCMC samples of the function $g$, and the right two panels show the posterior mean and variance of $\mathbf{y}^s(g)$.}
    \label{fig:inverse_result_single}
\end{figure}

Based on the constructed emulator, the Bayesian approach developed in Section \ref{sec:bayesianinverse} is then applied to infer the functional inverse given the observed response wave shown in the right panel of Figure \ref{fig:inverse_true}. Figure \ref{fig:inverse_demo} demonstrates the progression of the MCMC algorithm in evaluating the posterior distribution of the functional inverse $g$ as in \eqref{postg} and its corresponding predictive wave $\mathbf{y}^s(g)$ as in \eqref{postys}. As the algorithm progresses, both the sampled $g$ and the prediction of $\mathbf{y}^s(g)$ appear to be closer to the truth. The final posterior means of $g$ and $\mathbf{y}^s(g)$, demonstrated in the first and third panels of Figure \ref{fig:inverse_result_single}, appear to accurately recover the true function $g$ and its output $\mathbf{y}^s(g)$, which are shown in Figure \ref{fig:inverse_true}. Their posterior variances shown in the second and fourth panels of Figure \ref{fig:inverse_result_single}  allow for quantifying the uncertainties associated with the inverse recovery and prediction, which appear to be reasonably small.

To examine the performance, the proposed method is compared with existing alternatives based on the idea of KL expansion with finite truncation, which is a common approach in functional data analysis. 
%For the construction of the surrogate model $\mathbf{y}^s$, truncated KL expansion can be also applied to the functional inputs \citep{ramsay2005functional,yao2010functional,muller2013continuously,tan2019gaussian}, which is a common approach in the literature of functional data analysis where the problem is called \textit{scalar-on-function regression} \citep{ramsay2005functional, reiss2017methods}. Suggested by \cite{li2021gaussian}, the number of the truncated basis functions is selected that explains 95\% of the total variance.
The performance is evaluated by the root-mean-square error (RMSE) and the average proper score, which is the scoring rule  measuring the overall accuracy by taking into account both predictive mean and variance \citep{gneiting2007strictly}. Specifically, the proper score is defined as $-\left(\frac{y-\mu_P}{\sigma_P}\right)^2-\log\sigma^2_P,$
%$
%\text{proper score}=-\left(\frac{y-\mu_P}{\sigma_P}\right)^2-%\log\sigma^2_P,
%$
where $y$ is the true output, $\mu_P$ is the predictive mean, and $\sigma^2_P$ is the predictive variance. The larger score indicates better prediction performance. The reconstruction performance of $g$, including the RMSE and proper score, is computed based on the realizations of the true function $g(x_1,x_2)=1-\sin(x_2)$ and the posterior of $g$ on a set of grid points of size $101$ in the space $\Omega=[0,1]^2$.

Functional inputs are not only involved in constructing the emulator as discussed in Section 2.1, but also in the functional inverse developed in Section 2.2. To construct an emulator with functional inputs, the idea of truncated KL expansion is widely adopted in the literature of functional data analysis \citep{ramsay2005functional,reiss2017methods,yao2010functional,muller2013continuously,tan2019gaussian}. To recover the functional inverse,   
%The KL expansion can be applied to the correlation function in the GP prior \eqref{eq:GPprior} for functional inverse  \citep{li2021gaussian}, and it can also be applied to the functional input when constructing the surrogate model with functional inputs \citep{tan2019gaussian}. 
 \cite{li2021gaussian} propose to approximate the correlation function in the GP prior \eqref{eq:GPprior} by the expectation with respect to  $\boldsymbol{\eta}$, which can be written as  
$$
C(\mathbf{x},\mathbf{x}')=\int\Phi_{\boldsymbol{\eta}}(\mathbf{x},\mathbf{x}')\pi(\boldsymbol{\eta})\rm{d}\boldsymbol{\eta},
$$
and  KL expansion and truncation are then applied to obtain the first few leading eigenfunctions of $C(\mathbf{x},\mathbf{x}')$.
Although this approximation leads to fast computation, it cannot flexibly capture different levels of underlying smoothness. Additionally, it can introduce additional model bias caused by the truncation from basis expansion.

\begin{figure}[!h]
    \centering
    \includegraphics[width=\textwidth]{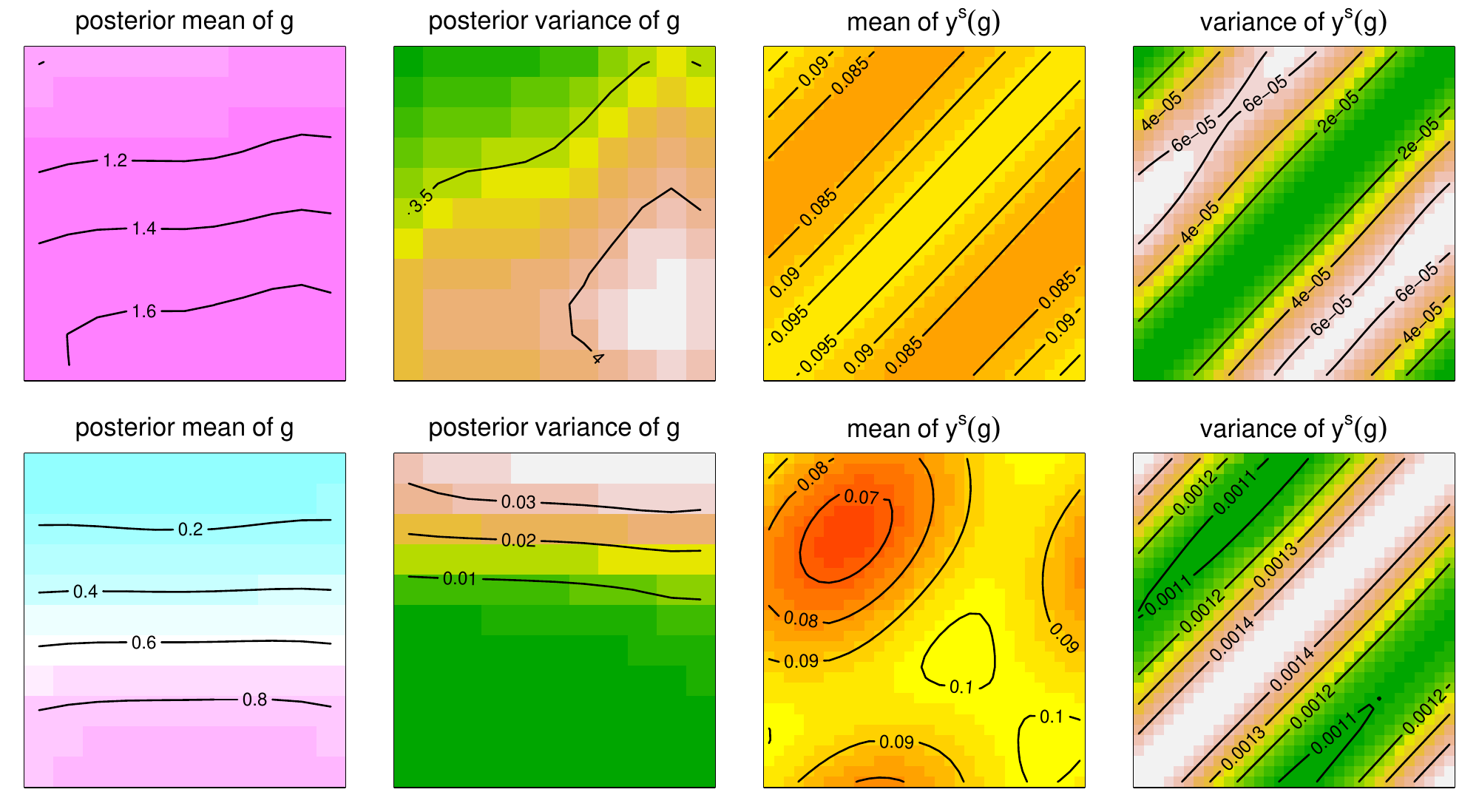}
    \caption{The results based on KL-expansion. The top row shows the results when KL-expansion is applied to both the emulator construction and functional inverse, which corresponds to the first row of Table \ref{tab:fidelity_comparison}. The bottom row shows the results when KL-expansion is applied to functional inverse, which corresponds to the second row of Table \ref{tab:fidelity_comparison}.}
    \label{fig:inverse_result_single_KL}
\end{figure}

Table \ref{tab:fidelity_comparison} presents the numerical comparison of the proposed method with two different scenarios using the idea of KL-expansion. One applies KL-expansion to both emulator construction and functional inverse (the first row in Table \ref{tab:fidelity_comparison}), and the other only applies it to functional inverse (the second row in Table \ref{tab:fidelity_comparison}). The reconstructed inverse and the corresponding far-field predictions are shown in Figure \ref{fig:inverse_result_single_KL}.
%applies to the inverse function $g$ and/or the training functional inputs $\{g_i\}^{10}_{i=1}$ for the construction of the surrogate model $\mathbf{y}^s(g)$. 
When the KL-expansion is applied to both emulator construction and inverse recovery (first row), the RMSE values for both  functional recovery of $g$ and output prediction of $\mathbf{y}^s(g)$ are relatively large. When the KL expansion is replaced by the proposed FIGP (the second row) for constructing the emulator, the RMSE is significantly reduced as compared to the first row. This observation is consistent with the results shown in Figure \ref{fig:inverse_result_single_KL}, where the second row outperforms the first one in both inverse recovery and far-field prediction.
This finding reveals the importance of preserving the functional input information through kernels as proposed in FIGP, as compared to KL-expansion where finite truncation introduces additional bias to the model.  The third row represents the proposed method, FIGP+GP, where FIGP is utilized for emulator construction and GP is used for estimating the functional inverse. The results demonstrate that the proposed method outperforms the other two alternatives in terms of RMSEs and proper scores. 
%When applying the expansion to the construction of the surrogate model (see the first row), the results of both the recovered $g$ and the reconstruction of $\mathbf{y}^s(g)$ exhibit poor performance. This is not surprising, as the emulation performance through truncated KL expansion has been shown to be inadequate in \cite{sung2022functional} due to insufficient number of finite basis functions. With the proposed FIGP emulator, the reconstruction performance of $\mathbf{y}^s(g)$ can be substantially improved  (see the second and third rows). When combined with the proposed GP for the inverse $g$, it yields superior performance in both the recovered $g$ and the reconstruction of $\mathbf{y}^s(g)$. 
%This verifies that a finite truncation of the basis functions can lead additional bias to the model \citep{sung2022functional}.

\begin{table}[t]
\begin{center}
\begin{tabular}{ C{3cm}|C{1.6cm}|C{1.6cm}|C{1.6cm}|C{1.6cm}|C{2cm}|C{1.6cm} } 
 \toprule
 &  &  & \multicolumn{2}{c|}{recovered $g$}  & \multicolumn{2}{c}{$\mathbf{y}^s(g)$} \\ 
  \midrule
&emulator & inverse & RMSE & Score & RMSE & Score\\
 \midrule
\multirow{3}{*}{FEM}&KL & KL & 0.815 &-1.493& 0.008572 & 8.73\\ 
  &FIGP & KL &0.054 &4.436& 0.000571 & 6.67\\ 
  &FIGP & GP &$\boldsymbol{0.028}$ &5.623& 0.000456 & 14.31\\ 
 \midrule
FEM + Born &FIGP &GP &  0.033 & $\boldsymbol{5.782}$ & $\boldsymbol{0.000243}$ & $\boldsymbol{15.857}$\\ 
 \bottomrule
\end{tabular}
\end{center}
    \caption{Performance of the recovered function $g$ and $\mathbf{y}^s(g)$ via single-fidelity simulations (FEM) and multi-fidelity simulations (FEM + Born). Better performances are highlighted as boldfaced.}
    \label{tab:fidelity_comparison}
\end{table}

\subsection{Inverse results with multi-fidelity simulation}\label{sec:resultmulti}
In this subsection, the Bayesian framework with the multi-fidelity surrogate modeling developed in Section \ref{sec:multisurrogate} is applied, which leverages the computational efficiency from the lower-fidelity simulations (Born approximation shown in Figure \ref{fig:realcase_real_born}) to enhance predictions for the high-fidelity simulations (FEM shown in Figure \ref{fig:realcase_real}). The functional-input GPs are applied as in \eqref{eq:multiFIGP}, and the LOOCVs are computed for each of the functional-input GPs with a linear and nonlinear kernel, the results of which are summarized in Table \ref{tab:loocv_multi}. It shows that the linear kernel is preferred for the functional-input GPs $\{h_l(g)\}^3_{l=1}$, while the nonlinear kernel is preferred for $\{\delta_l(g)\}^3_{l=1}$. The estimated autoregressive parameters are $\hat{\rho}_1=0.62,\hat{\rho}_2=1.35$, and $\hat{\rho}_3=1.35$.

\begin{table}[t]
\begin{center}
\begin{tabular}{ c|c|C{2.6cm}|C{2.6cm}|C{2.6cm} } 
 \toprule
 & Kernel  & $h_1(g)$ & $h_2(g)$& $h_3(g)$ \\ 
 \midrule
 \multirow{2}{*}{LOOCV} &  linear& $\boldsymbol{6.3\times 10^{-16}}$ & $\boldsymbol{4.1\times 10^{-16}}$ & $\boldsymbol{4.2\times 10^{-16}}$\\ 
 & nonlinear& $6.6\times 10^{-7}$ & $1.2\times 10^{-6}$ & $1.1\times 10^{-6}$\\ 
  \midrule
  & Kernel  & $\delta_1(g)$ & $\delta_2(g)$& $\delta_3(g)$\\ \midrule
 \multirow{2}{*}{LOOCV} &  linear& $7.5\times 10^{-5}$ & $7.6\times 10^{-18}$ & $9.5\times 10^{-20}$\\ 
 & nonlinear& $\boldsymbol{6.8\times 10^{-5}}$ & $\boldsymbol{7.3\times 10^{-18}}$ & $\boldsymbol{8.9\times 10^{-20}}$\\  
 \bottomrule
\end{tabular}
\end{center}
    \caption{The leave-one-out cross-validation errors (LOOCVs) for each of the functional-input GPs based on the multi-fidelity  simulations. The errors corresponding to the optimal kernel are boldfaced.}
    \label{tab:loocv_multi}
\end{table}

\begin{figure}[t]
    \centering
    \includegraphics[width=\textwidth]{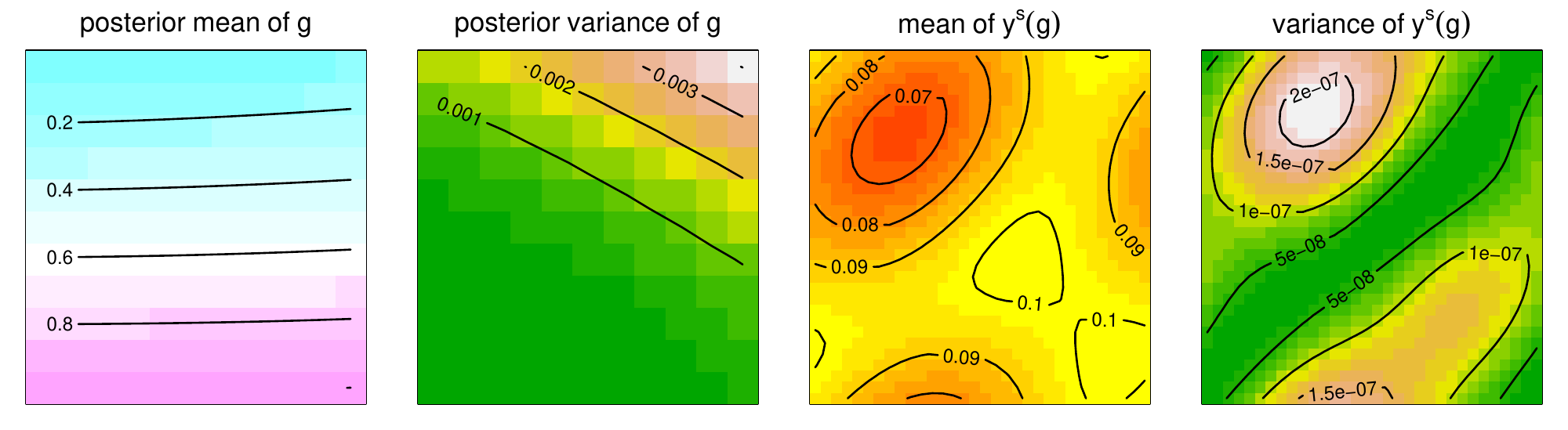}
    \caption{The inverse result based on the multi-fidelity simulator (FEM and Born approximation). The left two panels show the mean and variance of the MCMC samples of the function $g$, and the right two panels show the posterior mean and variance of $\mathbf{y}^s(g)$.}
    \label{fig:inverse_result_multi}
\end{figure}

The posterior means and variances of $g$ and $\mathbf{y}^s(g)$ are shown in Figure \ref{fig:inverse_result_multi}. It appears that both $g$ and $\mathbf{y}^s(g)$ are recovered fairly accurately with reasonably small uncertainties. The RMSE and average proper score are reported in the last row of Table \ref{tab:fidelity_comparison}. While it exhibits comparable performance to the single-fidelity simulator in reconstructing $g$ in terms of RMSE, utilizing the low-fidelity (Born approximation) simulations increases the proper score, which indicates the advantage of a smaller variation in the inverse reconstruction. Furthermore, integrating multi-fidelity simulations enhances
the reconstruction accuracy of $\mathbf{y}^s(g)$, which leads to a smaller RMSE and a better proper score. This observation agrees with the findings in computer experiment literature \citep{kennedy2000predicting,sung2022stacking}. 
To sum up, the proposed Bayesian inverse method can provide accurate recovery of functional input $g$ in both single and multi-fidelity emulators. %and the multi-fidelity emulator can enhance the reconstruction of the far-field pattern with the aid of the Born approximation simulator.

\section{Concluding Remarks}

The reconstruction of functional inverse in scattering problems is of increasing interest in science and engineering. In this article, a new Bayesian framework is introduced that includes a surrogate model, which accounts for functional inputs directly through kernel functions, and a hierarchical Bayesian procedure that infers functional inputs through the posterior distribution. The main contribution of the proposed method is the ability to prevent model bias caused by finite basis expansion, a common approach in functional data analysis, by employing a Gaussian process prior with linear and nonlinear kernels defined directly in the functional space. Furthermore, the proposed Bayesian approach integrates multi-fidelity simulations to enhance the  accuracy for  functional inverse reconstruction.

%utilizes a new surrogate model with functional inputs for efficient and accurate estimation of functional inputs in inverse scattering problems. The proposed approach overcomes limitations of traditional methods by employing a Gaussian process prior for the inverse solution without the need of finite basis expansion, and incorporating multifidelity simulations with functional inputs to enhance predictions for higher-fidelity simulations. The promising results obtained in our case study demonstrate the potential of this approach for advanced imaging and characterization techniques in fields such as medical imaging, geophysics, and materials science. 

%Our work establishes the groundwork for addressing inverse problems with functional inputs. 
%There are potential extensions based on the proposed Bayesian method. 

Besides MCMC discussed in the paper, it is of our interest to explore alternatives including variational Bayesian inference \citep{jordan1999introduction,wainwright2008graphical,tran2015variational,hensman2015scalable, blei2017variational}, which has the potential to speed up the computation and achieve a better scalability, particularly for large datasets.
%First, while the MCMC method developed herein shows promise and efficiency, it would be worthwhile to explore the use of variational Bayesian inference techniques \citep{jordan1999introduction,wainwright2008graphical,blei2017variational}, which could lead to faster computations and improved scalability, particularly for large datasets. Recent studies on variational GPs, such as \cite{tran2015variational} and \cite{hensman2015scalable}, could be relevant in this context. 
Another interesting ongoing research extended from the proposed method is the study of experimental designs in a  functional space. Despite numerous studies on optimal design in computer experiments, the developments for functional inputs are scarce. In particular, the widely used space-filling designs in computer experiments are based on the distance measures defined in the Euclidean space, which are not directly applicable to a functional space. Furthermore, the development of efficient experimental designs for multi-fidelity simulations in a functional space is also of interest in practice.

%Additionally, despite the successful application of our Bayesian framework, it would be valuable to explore a frequentist version for inverse scattering problems. This could involve replacing the Gaussian process prior with a function that resides in a Reproducing Kernel Hilbert Space (RKHS), and considering a loss function with a penalized RKHS norm, which could provide some theoretical guarantees. We anticipate that this approach could yield interesting theoretical guarantees, as demonstrated in recent developments along this line of research, such as in \cite{tuo2021reproducing}.  The extension to develop a frequentist approach for functional outputs presents an interesting and important avenue for our future research.

\vspace{0.5cm}
\noindent\textbf{Supplemental Materials}
The \texttt{R} code for reproducing the results in Section \ref{sec:application} is provided in Supplemental Materials.

\section*{Appendix}
\begin{appendix}
\section{Sampler Details}\label{sec:suppsampler}
We introduce the sampler for the distribution $\pi(\mathbf{g}_N,\sigma^2_e,\boldsymbol{\eta},\tau^2_g|\mathbf{y}^p)$ via Gibbs sampling, which is drawn iteratively from 
\begin{align}
&\pi(\mathbf{g}_N|\sigma^2_e,\boldsymbol{\eta},\tau^2_g,\mathbf{y}^p),\label{eq:supp1}\\
&\pi(\sigma^2_e|\mathbf{g}_N,\boldsymbol{\eta},\tau^2_g,\mathbf{y}^p),\label{eq:supp2}\\
&\pi(\boldsymbol{\eta}|\sigma^2_e,\mathbf{g}_N,\tau^2_g,\mathbf{y}^p),\label{eq:supp3}\\
&\pi(\tau^2_g|\sigma^2_e,\mathbf{g}_N,\boldsymbol{\eta},\mathbf{y}^p).\label{eq:supp4}
\end{align}
The conditional distribution \eqref{eq:supp1} can be drawn via Metropolis-Hastings algorithm. We use a multivariate normal distribution as a proposal distribution that draws the proposed sample  $\mathbf{g}'_N$ by 
\begin{equation*}\label{eq:proposal}
\mathbf{g}'_N=c_g\tau_g\boldsymbol{\Phi}_{\boldsymbol{\eta}}^{1/2}\mathbf{Z}_{N}+ \mathbf{g}_N,    
\end{equation*}
where
$\mathbf{Z}_{N}\sim\mathcal{N}_{N}(0,\mathbf{I}_{N})$ and $\mathbf{I}_{N}$ is an identity matrix of size $N$, and $c_g>0$ is a small constant which can be adaptively determined by monitoring the acceptance rate. We accept $\mathbf{g}^{(k+1)}=\mathbf{g}_N'$ with the probability
$$
\min\left\{1,\frac{\pi(\mathbf{g}_N^{(k+1)},(\sigma^2_e)^{(k)},\boldsymbol{\eta}^{(k)},(\tau^2_g)^{(k)}|\mathbf{y}^p)}{\pi(\mathbf{g}_N^{(k)},(\sigma^2_e)^{(k)},\boldsymbol{\eta}^{(k)},(\tau^2_g)^{(k)}|\mathbf{y}^p)}\right\}.
$$
where $\pi(\mathbf{g}_N,\sigma^2_e,\boldsymbol{\eta},\tau^2_g|\mathbf{y}^p)$ is the distribution as in \eqref{eq:jointdist}, and the superscript $k+1$ and $k$ indicate the $(k+1)$-th and $k$-th iterations, respectively. 

The conditional distribution \eqref{eq:supp2} also can be drawn via Metropolis-Hastings algorithm, where the proposal is drawn by 
$$\log((\sigma_e^2)')=c_sz+\log(\sigma_e^2),$$
where $z$ is a standard normal distribution and $c_s$ is a small constant. We accept $(\sigma_e^2)^{(k+1)}=(\sigma_e^2)'$ with the probability
$$
\min\left\{1,\frac{\pi(\mathbf{g}^{(k)}_N,(\sigma_e^2)^{(k+1)},\boldsymbol{\eta}^{(k)},(\tau^2_g)^{(k)}|\mathbf{y}^p)}{\pi(\mathbf{g}^{(k)}_N,(\sigma_e^2)^{(k)},\boldsymbol{\eta}^{(k)},(\tau^2_g)^{(k)}|\mathbf{y}^p)}\right\}.
$$
The parameters $\boldsymbol{\eta}^{(k+1)}$ can be drawn in a similar fashion from \eqref{eq:supp3}. The sample of $(\tau_g^2)^{(k+1)}$ can be also drawn by its posterior distribution \eqref{eq:supp4}, which is an inverse gamma distribution with the shape parameter $a_2+N/2$ and the rate parameter 
$$
b_2+\frac{1}{2}\left((\mathbf{g}_N^{(k)})^T\boldsymbol{\Phi}_{\boldsymbol{\eta}^{(k)}}^{-1}\mathbf{g}_N^{(k)}\right).
$$

\end{appendix}

\bibliography{Inverse_references}

\iffalse
\def\spacingset#1{\renewcommand{\baselinestretch}%
{#1}\small\normalsize} \spacingset{1.3}

\newpage
\setcounter{page}{1}
\bigskip
\bigskip
\bigskip
\begin{center}
{\Large\bf Supplementary Materials for ``Advancing Inverse Scattering with Surrogate Modeling and Bayesian Inference for Functional Inputs''}
\end{center}
\medskip

\setcounter{section}{0}
\setcounter{equation}{0}
\def\theequation{S\arabic{section}.\arabic{equation}}
\def\thesection{S\arabic{section}}

\section{Sampler Details}
\fi
\end{document}